# Living Structure Down to Earth and Up to Heaven: Christopher Alexander

Bin Jiang

Faculty of Engineering and Sustainable Development, Division of GIScience
University of Gävle, SE-801 76 Gävle, Sweden
Email: bin.jiang@hig.se

(Draft: July 2019, Revision: July and August 2019)

> *"All of my life I've spent trying to learn how to produce living structure in the world. That means towns, streets, buildings, rooms, gardens, places which are themselves living or alive… depending on who you talk to, they'd say, 'Well, this stuff Alexander's been discovering is a lot of nonsense. There is no such thing as objectivity about life or quality.' … They are simply mistaken."*
>
> Christopher Alexander (1999)

**Abstract:**
Discovered by Christopher Alexander, living structure is a physical phenomenon, through which the quality of the built environment or artifacts can be judged objectively. It has two distinguishing properties just like a tree: "far more small things than large ones" across all scales from the smallest to the largest, and "more or less similar things" on each scale. As a physical phenomenon, and mathematical concept, living structure is essentially empirical, discovered and developed from miniscule observation in nature- and human-made things, and it affects our daily lives in some practical ways, such as where to put a table or a flower vase in a room, helping us to make beautiful things and environments. Living structure is not only empirical, but also philosophical and visionary, enabling us to see the world and space in more meaningful ways. This paper is intended to defend living structure as a physical phenomenon, and a mathematical concept, clarifying some common questions and misgivings surrounding Alexander's design thoughts, such as the objective or structural nature of beauty, building styles advocated by Alexander, and mysterious nature of his concepts. For this purpose, we first illustrate living structure – essentially organized complexity, as advocated by the late Jane Jacobs (1916–2006) – that is governed by two fundamental laws (scaling law and Tobler's law), and generated in some step by step fashion by two design principles (differentiation and adaptation) through the 15 structural properties. We then verify why living structure is primarily empirical, drawing evidence from Alexander's own work, as well as our case studies applied to the Earth's surface including cities, streets, and buildings, and two logos. Before reaching conclusions, we concentrate on the most mysterious part of Alexander's work – the luminous ground or the hypothesized "I" – as a substance that pervasively exists everywhere, in space and matter including our bodies, in order to make better sense of living structure in our minds.

## 1. Introduction

If the life's work of Alexander (2002–2005) – *The Nature of Order* – had to be summarized in one word, "beauty", "life" and "wholeness" would be the three top candidates. If allowed two words, it would be "living structure". What do these terms really refer to? Instead of getting into their detailed meanings (see Section 2), let us use an analogue to clarify them first. If wholeness were compared to temperature, then beauty or life would be like the feeling of warmness or coldness. The higher the temperature, the warmer one feels, and the lower the temperature, the colder one feels. The higher the wholeness, the more beautiful or the more life one feels; the lower the wholeness, the less beautiful or the less life one feels. Therefore, a thing or structure that exhibits a high degree of wholeness is called a living structure. Opposite to living structure is non-living (or dead) structure. There is a wide range between the living and the dead, so living is always to some degree or other, just as the feeling of warmness relates to a range of temperatures. Living structure is what Alexander (2002–2005) discovered and further pursued, and it is independent of any style or culture from for example Indonesia, Japan, Russia, Africa, Turkey, Iran, India, or China. Having said that, Alexander has no particular style of buildings, contrary to what his rivals or critics tend to think. Or to put it different, living structure is a living style, just as nature itself, being able to trigger a sense of belonging or well-being or healing in people who are exposed to it. To know whether a thing or space exhibits living structure, one can simply examine whether it possesses "far more smalls than larges" across all scales ranging from the smallest to the largest. For example, at the multiple levels of scale or in a recursive manner – an entire tree, its branches, and its leaves (in terms of the detailed texture) – there are always "far more smalls than larges". Therefore, a tree is beautiful or alive structurally, regardless of whether it is alive biologically.

Based on the notion of "far more smalls than larges", a simple shape that lacks of detailed smaller structures is neither beautiful nor alive. This is for the same reason why sans-serif fonts are less beautiful or less alive than serif ones. For example, the font "I" (when shown as a sans-serif) is not a living structure (one vertical line only) without "far more smalls than larges", whereas the font "I"(when shown as a serif) is a relatively living structure (one vertical line and two little bars) with "far more smalls than larges". The difference between the non-living and living fonts may be hardly sensed when the two fonts are too small, in particular when the letter's meaning is focused on. As a matter of fact, serif fonts in general are objectively more beautiful than sans-serif ones. It is based on this kind of structural fact – actually the phenomenon of living structure – that Alexander (2002–2005) established a scientific foundation of architecture. Unfortunately, the phenomenon of living structure has not yet been well accepted by the scientific community as a fact, but been sidelined as a human taste or personal preferences. This situation constitutes a major motivation of this paper. Human history is full of many great builders or architects who made great buildings, but few of them really made it clear – or even intended to think about – how to make great buildings and why the great buildings are great.

As an architect who was initially trained in science, Alexander wanted to make beautiful buildings, and he wanted to know in particular why beautiful buildings are beautiful. His classic work on the pattern language (Alexander et al. 1977) is widely read by ordinary people looking to make beautiful rooms, houses and gardens, and to facilitate their daily lives, for example, where to put a lamp or a flower vase, and how to lay a table cloth. His design thoughts are therefore very practical – down-to-earth – and his research is essentially empirical. On the other hand, his research is deeply philosophical and visionary, up to heaven, touching the fundamental issues of what the universe constitutes, and where our consciousness comes from. He conceived and developed – from the phenomenon of living structure – a third view of space, and a new cosmology in which we human beings – not only the body but also the mind – are part of the universe (Alexander 2002–2005, Volume 4, c.f. Section 5 for a more detailed discussion). The third view of space states that space is neither lifeless nor neutral, but a living structure capable of being more living or less living. This new view of space sets a clear difference from two traditional views of space: absolute space by Isaac Newton (1642–1727) and relational space by Gottfried Wilhelm Leibniz (1646–1716), both of which are framed under the mechanistic world view of Descartes (1637, 1954). This new view of space constitutes part of the new cosmology that unifies the physical world and our inner world as a coherent whole.

Despite a large body of literature on or inspired by Alexander's work (e.g., Gabriel 1998, Salingaros 2006, Quillien 2008, Jiang and Sui 2014, Leitner 2015, Wania 2016, Mehaffy 2017, Guttmann et al. 2019, Jiang 2019a), living structure has not yet been well recognized as a physical phenomenon or mathematical concept, for people to understand the objective or structural nature of beauty. This paper is an attempt to fill this gap, by setting up a dialogue with those who are skeptical about Alexander's design thoughts. It is intended to clarify some doubts in order for skeptics to understand three main points. First, the essence of beauty is structural or objective, lying in the notion of "far more smalls than larges", which accounts for a majority of our sense of feeling on beauty. There is a clear sign that beauty is beginning to be accepted as an objective concept in the literature of philosophy (Scruton 2009). Second, the phenomenon of living structure is universal and pervasive, not only in nature but also in what we made and built across all cultures, ethics, and religions, involving ancient buildings and cities, as well as ancient carpets and other artifacts. Thus, there is no so-called Alexander's style of architecture; if there is, it is the living structure (just as nature itself), which is able to trigger a sense of beauty or life in the human mind. Third, there is no mystery at all regarding the *"quality without a name"* (Alexander 1979), which is actually living structure, yet the mystery of a non-material world view remains.

The remainder of this paper is structured as follows. Section 2 introduces and illustrates the living structure as a physical phenomenon, using a sketch by Alexander, in terms of its governing laws (scaling law and Tobler's law), its design principles (differentiation and adaptation), and its 15 structural properties (Table 1). Section 3 argues why living structure is scientific or empirical by drawing evidence from Alexander's own works such as the pattern language. Section 4 further presents case studies to demonstrate that living structure is objective or structural rather than just a matter of opinion. Section 5 discusses the metaphysical aspects in order to make better sense of the living structure in terms of why living structure evokes a sense of beauty or life in our minds. The paper concludes with a few remarks and suggestions for future work.

**2. Living structure: Its governing laws and design principles, and 15 structural properties**
The four terms mentioned at the outset of this paper can be placed into two categories − wholeness and living structure in the first group, and beauty and life in the second group − representing the outer and inner worlds, respectively. The central concept among these four is wholeness, which can be defined mathematically (Alexander 2002–2005, Jiang 2015b). It exists pervasively in our surroundings; in an ornament, in a room, in a building, in a garden, and in a city. It was previously referred to by Alexander as the *"quality without a name"*: *"a central quality which is the root criterion of life and spirit in man, a town, a building, or a wilderness. This quality is objective and precise, but it cannot be named"* (Alexander 1979). The term wholeness is also a key concept in Gestalt psychology (Köhler 1947), in quantum physics (Bohm 1980), and in many other religious and philosophical contexts. Semantically, there may be some overlap across these different fields, but Alexander's wholeness is unique with its distinguishing features. It is not a static structure, but a dynamic process, through which living structure emerges. In the next part of this section, we will use a sketch by Alexander (2002–2005) (6a in Figure 1) to introduce and illustrate living structure or wholeness.

Table 1: The 15 structural properties of wholeness

| Levels of scale | Good shape | Roughness |
|---|---|---|
| Strong centers | Local symmetries | Echoes |
| Thick boundaries | Deep interlock and ambiguity | The void |
| Alternating repetition | Contrast | Simplicity and inner calm |
| Positive space | Gradients | Not separateness |

The sketch shows the evolution of a living structure, demonstrating many of the 15 properties (Table 1). It consists of at least 19 different sized mutually overlapping, nested shapes or centers in Alexander's terms, namely the four outmost black dots, the square, the big circle, the eight tiny triangles, the four small circles, and the tiny dot in the middle. Among many of the other properties, the first property of the levels of scale is the most distinguishing one. The sketch is with six levels of scale, indicated by the six colors with red being the highest, blues being the lowest, and other colors in between (6b in Figure 1). It should be noted the six levels of scale is not in terms of their sizes, but the supports they receive. Overall, there are "far more blues than red", and some in between the blues and the red in the colored sketch, the spectral coloring in terms of the degree of wholeness of individual centers. For example, the tiny center in the middle has the highest degree because it receives many supports from other centers. The notion of "far more smalls than larges" reflects the very first property of living structure, namely levels of scale (Table 1), or "scaling hierarchy" or scaling law (Jiang 2015c). As a reminder, the scaling hierarchy of "far more smalls than larges" should be – more correctly – understood in a recursive manner, implying that "far more smalls than larges" for a whole, for a sub-whole, and a sub-whole of sub-wholes and so on (c.f. the above tree example). The scaling hierarchy of "far more smalls than larges" recurs multiple times rather than just once, except for some simple cases like font "I". In the living structure, the recurring happens five times (steps 2–6 in Figure 1), leading to six hierarchical levels. The living structure can be said, more precisely, to be evolved, which implies that centers are well adapted to each other as a coherent whole. The living structure is not simply an assembly of pre-existing components. In this regard, the less-living structure (6c in Figure 1) is indeed an assembly of pre-existing units.

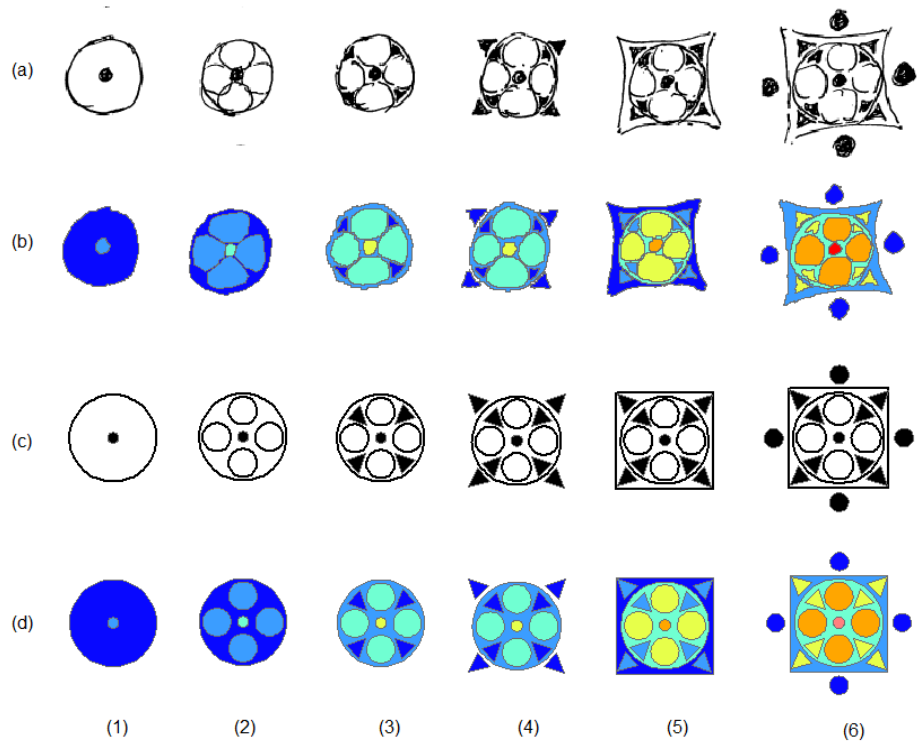

Figure 1: (Color online) A living (6a and 6b) versus a less-living structure (6c and 6d)
(Note: The living structure is evolved as a product of living process, while the less-living structure comes from a simple drawing. The living structure is well differentiated and well adapted, not only among those in the figure, but also among those in the ground, whereas the less-living structure looks like an assembly from pre-determined pieces, which are not well adapted to each other. For example, the four dots outside the square in 6c and 6d are less integrated into the whole than those in 6a and 6b. The coloring indicates the degree of wholeness, with blues showing the lowest, red showing the highest, and other colors in between the lowest and highest.)

The less-living structure (6c and 6d in Figure 1) looks smooth, glassy, and uniform, but it exhibits a lower degree of wholeness. There are several reasons for this, a few of which are highlighted here. First, the less-living structure is created (at once) by assembling rather than generated (step by step) by adaptive design. The lack of adaptation can be clearly seen from figure and ground relationship (Rubin 1921), or the fact that the space between these geometric shapes is not well shaped, or not convex like ripening corn, *"each kernel swelling until it meets the others, each one having its own positive shape caused by its growth as a cell from the inside"* (Alexander 2002–2005). Second, the four dots outside the square of the less-living structure are less integrated into the whole inside the square. This is because the space outside of square is fully open and therefore lacks a sense of belonging for the four dots, as strikingly shown in the living structure. Third, the less-living structure misses the property of roughness. Interested readers can compare against the 15 structural properties (Table 1) to learn why one structure is more living than the other: the more structural properties, the more living a structure is.

There are two fundamental laws of living structure, scaling law and Tobler's law, which also underlie the 15 structural properties (Table 1). The first structural property (levels of scale) reflects scaling law, as elaborated above, while the remaining properties are largely a reflection of Tobler's law (Tobler 1970), probably except for the property of "not separateness". Tobler's law, which is commonly called the first law of geography, states that *"everything is related to everything else, but near things are more related than distant things"*. Essentially, Tobler's law is complementary to – rather than contradictory to – scaling law (Table 2), indicating that on each scale, centers are "more or less similar". It is important to stress that centers that are "more or less similar" are more beautiful or more alive than centers that are precisely the same. For example, in the living structure, the four small circles are "more or less similar", and eight triangles are "more or less similar", so they are much more living than if they were precisely the same in size, as shown in the less-living structure. With the perfectly drawn shapes, some of the 15 structural properties are no longer available, such as positive space and roughness, which can be phrased as *"the perfection of imperfection"* (Junker 1991). These two properties are the most important for naturally evolved things, such as cell structures and maize grains. On the surface, naturally evolving things may look rather rough or irregular, yet they tend to exhibit the essence of natural beauty. As for Tobler's law or the notion of "more or less similar" on each scale, we can add another example: a coastline with the same degree of complexity as the Koch curve (Koch 1904), with which things (or segments) are precisely the same at each of scales such as 1/3, 1/9, 1/27 and so on. The coastline at each of its scales exhibits the property of "more or less similar" segments rather than precisely the same ones, so the coastline is more natural, more beautiful, or more living than the Koch curve.

Table 2: Scaling law and Tobler's law of living structure
(Note: These two laws are complementary of – rather than contradictory to – each other and they recur at different levels of scale of living structure.)

| Scaling law | Tobler's law |
|---|---|
| far more small things than large ones | more or less similar things |
| across all scales | available on one scale |
| without an average scale (Pareto distribution) | with an average scale (Gauss distribution) |
| long tailed | short tailed |
| interdependence or spatial heterogeneity | spatial dependence or homogeneity |
| disproportion (80/20) | proportion (50/50) |
| complexity | simplicity |
| non-equilibrium | equilibrium |

From a design or dynamic point of view, a space, or structure is continuously differentiated toward scaling hierarchy of "far more smalls than larges". Actually, these two laws are largely statistical, which does not guarantee living structure. To make the structure really living or beautiful, we must consider geometric aspects. This is the idea of adaptation: on each level, things should be "more or less similar", or nearby things should be "more or less similar". Note that "nearby" is usually referred to in a geometrical distance, but a topological distance is better in many instances. For example, my neighbor is defined within a certain geometric distance, but an airport's neighbor is better defined in terms of topological distance of flight connections. My neighbor's house should look "more or less similar" (in size and shape) to my house, whereas Heathrow Airport should look "more or less similar" (in size or capacity) to the Paris Charles De Gaulles Airport rather than the Gatwick Airport, because there is no flight between the Heathrow and the Gatwick. Along these two laws, there are two design principles: differentiation and adaptation. The living structure in Figure 1 is continuously differentiated to reach the status of living structure.

It is important to realize – as Alexander noted repeatedly – that the evolution process is not simply about adding new centers; more correctly, centers are induced by the wholeness. In other words, it is incorrect to say a whole comes from parts, or a whole consists of parts; it is the wholeness that induces centers to generate a coherent whole. It is incorrect to say a flower consists of petals; it is the flower as a whole that induces petals. Another design principle is adaptation. On each level of scale, saying that things are "more or less similar" implies things are adapted to each other. Again, this notion of "more or less similar" things should really be understood literally. If things are exactly the same, it tends to generate a structure that is less living or less beautiful; see also examples mentioned above about the coast line versus the Koch curve. It should be noted that adaptation could imply things adapted across scales. This is again a good example of Alexander's observation (see more in Section 3). Alexander found that, across levels of scale, the scaling ratio should be between 2 and 3; otherwise structure would look less living (see Figure 4 in Section 3). It is in this sense that Alexander's living geometry generally surpasses fractal geometry (Mandelbrot 1983). Fractal geometry hardly cares about whether the generated pattern is beautiful or not, and it only cares about automation of some structure.

### 3. A commonsense and humane approach to architecture
The phenomenon of living structure exists not only in human-made or -built things, but also in nature. Alexander's approach to architecture is very much commonsense and humane. More importantly, he wanted to be inspired by nature and to make sure that what he observed from what humans built or made also applied to nature. For example, the 15 properties of living structure are pervasively seen not only in the built environment, but also in nature. In this section, we draw evidence from Alexander's earlier works to learn why living structure is scientific and empirical, and why this is a correct way of conducting science and art.

Alexander first described the idea of living structures in a corner of an English country garden, where a peach tree grew against a wall:

*"The wall runs east to west; the peach tree grows flat against the southern side. The sun shines on the tree and, as it warms the bricks behind the tree, the warm bricks themselves warm the peaches on the tree. It has a slightly dozy quality. The tree, carefully tied to grow flat against the wall; warming the bricks; the peaches growing in the sun; the wild grass growing around the roots of the tree, in the angle where the earth and roots and wall all meet."* (Alexander 1979)

In this living structure of the garden corner, there are many interconnected living centers, such as the wall, the peach tree, the sun, the bricks, the wild grass, the roots of the tree, and even the garden. This is a very good example of Alexander's miniscule observations on nature and on our surroundings.

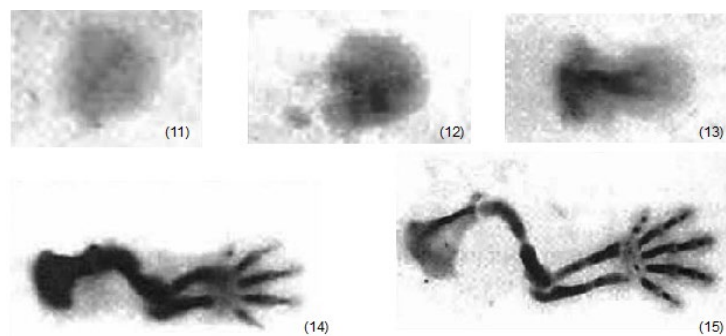

Figure 2: The mouse foot as a living structure emerging from day 11 to 15 (Alexander 2005)
(Note: Every step is based on the previous step for enhancing the degree of wholeness. Apparently, the number of centers induced is increased in the course of growth from day 13 to 15.)

Considering another example of embryogenesis, a growing mouse foot is a living structure that comes from continuous differentiation and adaptation (Figure 2, Alexander 2005). In the course of the step-by-step development of the five days, many of the 15 structural properties can be observed, such as strong centers, thick boundaries, gradients, levels of scale, contrast, local symmetries, and finally, good shape of the whole. Alexander started his research on architecture – nearly from scratch – not only from traditional buildings and cities, but also from ancient artifacts such as carpets. Two of his books have accurately documented his miniscule observations: one on built environment (Alexander et al. 1977) – part of the trilogy with Alexander et al. (1975) and Alexander (1979) – and the other on carpets (Alexander 1993). His dream was to build beautiful buildings and cities, sharing the same order or beauty – or spirit – of nature, and his understanding of the kind of natural beauty that exists in deep structure rather than on the surface, such as thermal comfort, energy saving or illumination of surfaces. Nowadays, so-called "green" buildings are not really green according to Alexander, and *"a world built according to the present sustainable paradigm, the technical sustainability paradigm, would be quite a horrible place"*. Instead, living structures *"represent true sustainability, they sustain the heart, and sustain the soul. They sustain the humanness of the person, and they sustain the Earth"* (Alexander 2004). When nature and the built environment are treated as one, when the physical world and our human's inner world are treated as one, and when the precious Earth's surface and our surroundings are treated as our garden, we will be able to reach the true sense of sustainability.

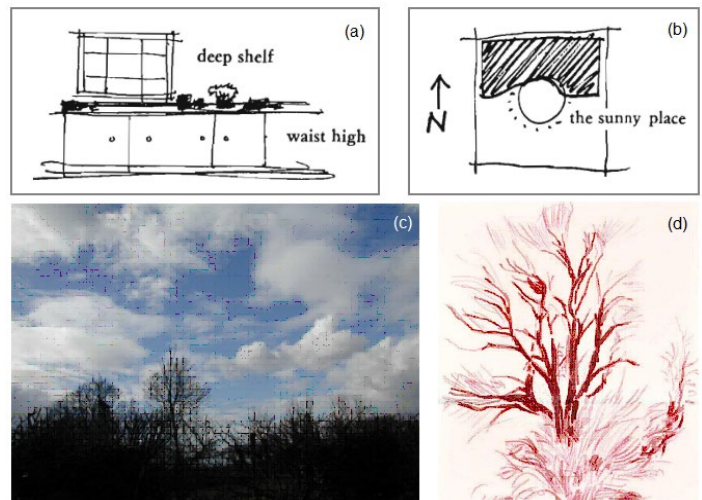

Figure 3: (Color online) Alexander's miniscule observation on architecture and nature
(Note: (a) a waist-high shelf, (b) a sunbeam coming into a room, (c) the interlocking blue sky and white clouds, and (d) positive space generated between a tree's branches (Alexander et al. 1977, Alexander 2005).)

Back in the 1970s, when he was granted a project by National Institute of Mental Health in the United States to investigate on relationship between the built environment and human well-being, Alexander and his colleagues started learning what works and what does not. To a great extent, traditional buildings and artifacts were his great teacher. His research is not limited to a certain type, but all types of building – across all cultures and countries – that people know, deep in their heart, are universally beautiful or alive. For example, he noted that a waist-high shelf can be very convenient for people to leave things while searching for keys (Figure 3a), whereas a room with a sunbeam is more uplifting (Figure 3b). Thus, both the shelf and sunbeam are supportive to human well-being. It is these kinds of materials that constitute the major content of pattern language (Alexander et al. 1977).

Alexander describes his early work on pattern language as follows:

*"To get my feet on the ground, and to have something solid that I could be sure of, I started by examining the smallest particles of functional effect, that I could discern in buildings, with small and sometimes barely significant aspects of the ways that buildings affect people. My purpose in doing this, was to focus on the smallest particles of fact that I could be certain of: something that was extraordinarily difficult when faced with the porridge of mush that then passed for architectural theory. In the early years my studies were based on the most ordinary, miniscule observations about usefulness and the effect of buildings on the people who lived in them, always keeping the observations modest, reliable -- small enough and solid enough so that I could be sure that they were true.*

*At first I included very small particulars of functional effect in any matter that actually made a practical difference to daily life… a shelf besides the door where one could put a packet down while searching for ones keys, for instance, or the possibility of a sunbeam coming into a room and falling on the floor.*

*But I quickly realized that some of these details were very much more significant than others. Those like the first (the shelf) tended to be pedestrian, even though useful; while those like the second (the sunbeam) were more uplifting, and clearly mattered more in some obvious but profound sense. I began to focus on those miniscule points which mattered more, in the sense of the second example. Gradually, then, I was able to pave the way to the possibility of seeing how buildings support human well-being – not so much mechanical or material well-being, but rather the emotional well-being that makes a person feel deeply comfortable in himself. And as I studied these small effects carefully, gradually I was led to a conception of wholeness, wellness, and spiritual support that might, under ideal circumstances, be present between buildings and human beings."* (Alexander 2007a)

Built on this earlier work of pattern language, Alexander realized that there are some structural aspects – the 15 properties – that are the most fundamental to human well-being. For example, the beauty of the blue sky and clouds comes from the property of positive space (Figure 3c); not only the white clouds but also the blue sky are well shaped. The same positive space appears between a tree's branches (as Alexander sketched himself in Figure 3d). This property of positive space is particularly important for urban environments. It implies that not only buildings, but also the space between buildings, should be well shaped like convex spaces, as we discussed with the living versus less-living structures presented in Figure 1.

Goodness of space is a matter of fact rather than opinion or personal preferences, and good space has a healing effect, sustaining and promoting health. For example, two spaces that are either too close or too open are transformed into positive spaces to which people can develop a sense of belonging (Figure 4 a and b). This sense of belonging further triggers human well-being, security, or safety. Well-being or comfort provided by environments or space is an important factor for human healing. In this regard, Ulrich (1984), an architect, found that the view from a window may influence a patient's recovery from surgery; that is, natural scenes are better than urban scenes for post-surgery recovery. Taylor (2006), a physicist, found that fractal patterns, if they are living structures, in nature and art have stress-reducing effect on people. Because both natural scenes and living structures are living rather than non-living, I conjecture that, essentially, it is living structures that have healing effects on people.

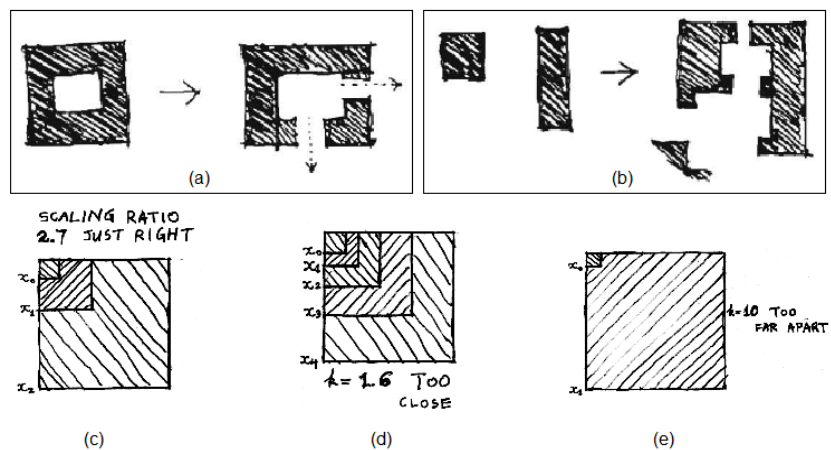

Figure 4: Good or bad as a matter of fact rather than an opinion
(Note: A bad space – (a) too close or (b) too open – is transformed into a good space, creating a sense of belonging (Alexander et al. 1977). The scaling ratio is between 2 and 3, as observed by Alexander (2002–2005), tends to lead to a good design or structure, while the scaling ratios of 1.6 (d) or 10 (e) are either too close or too far apart (Salingaros 2006).)

Alexander made another important discovery related to the property of levels of scale or scaling hierarchy. He found that the scaling ratio between two consecutive scales should be between 2 and 3; having it too close or too far apart would reduce the goodness or coherence of space or structure (Figure 4c, d and d). The major difference between fractal geometry (Mandelbrot 1983) and Alexander's work – or living geometry (Alexander et al. 2012) – is that the former is largely for understanding complexity of fractals. Although fractal geometry is able to create artificial fractals, it often ends with *"pretty pictures, pretty useless"* (Mandelbrot 1983). By contrast, living geometry aims not only to understand complexity, but also to create organized complexity, or beautiful or living structures, which are able to trigger a sense of beauty or life. The creation or the making of living structures is what makes Alexander (2003) differ from other pioneers in complexity science.

The following quote shows how Alexander explained the property of the levels of scale or scaling hierarchy pervasively seen in nature and in what we make and build, and how it triggers the feeling of life in our hearts:

*"I would like to summarize the content of this new kind of empirical complex in the following way. In any part of what we call nature, or any part of a building, we see, at many levels of scale, coherent entities or centers, nested in each other, and overlapping each other. These coherent entities all have, in varying degree, some quality of "life."*

*For any given center, this quality of life comes about as a result of cooperation between the other living centers at several scales, which surround it, which contain it, and which appear within it. The degree of life any one center has, depends directly on the degree of life that is in its associated centers at these different scales. In short, I had identified a kind of wholeness: in which the life of any given entity depended on the extent to which that entity had unfolded from the wholeness."* (Alexander 2007a)

In addition to what is summarized above, there are many other empirical findings (Alexander 2007b) that support living structure as a physical phenomenon, as well as a well-defined mathematical concept. In following section, we will carry out some case studies to support living structure is not only a phenomenon and concept, but also can be used to objectively judge quality of things.

**4. Case studies on living structure**
Digital technology, particularly geographic information systems, now provides enormous data about the Earth's surface, about cities, buildings, and about artifacts for revealing living structure in our surroundings. This section reports several case studies for revealing ubiquitous living structure. To make this paper more readable, we do not use the mathematical model of wholeness for computing the degree of wholeness (Jiang 2015b, Jiang 2016). Instead, the case studies do no more than count the number of centers, and compute scaling hierarchy of "far more small centers than large ones". Previous studies have illustrated that this simple way of computing degree of wholeness is good enough in particular for comparison purposes (Jiang 2018, Jiang and Ren 2018). Anyone who is able to count can easily follow the case studies. Before the case studies, it is necessary to first introduce head/tail breaks (Jiang 2013, 2015a), which helps compute the scaling hierarchy of "far more smalls than larges". The scaling hierarchy is visualized by a series of spectrum colors ranging from blue for the lowest to red for the highest; the more colors, the more levels of scale, the more beautiful or living. Through the coloring, the notion of "far more smalls than larges" is equivalent to "far more blues than reds"; see Figure 5d, 5e, 6, 7 and 8 for the coloring.

**4.1 Head/tail breaks for calculating scaling hierarchy**
Given a dataset with a heavy tailed distribution or with "far more smalls than larges", head/tail breaks can help obtain the inherent scaling hierarchy by recursively breaking the dataset into two parts (the head and the tail) around the mean (Jiang 2013, 2015a). Those greater than the mean are called the head, and those less than the mean are called the tail. To illustrate the head/tail breaks, consider the 10 numbers – 1, 1/2, 1/3, …, and 1/10 – that exactly follow Zipf's law (1949) as a working example. These 10 numbers are already ranked from the biggest to the smallest. Their mean is ~0.29, which partitions the 10 numbers into two groups: the biggest three as the head, and the smallest seven as the tail. The mean of the biggest three is ~0.61, which further partitions the largest three into two groups again: the biggest 1 as the head, and the smallest two 1/2 and 1/3 as the tail. The notion of "far more smalls than larges" recurs twice, so the scaling hierarchy is three (or, in other words, three levels of scale). In general, the head/tail breaks is formatted as a recursive function as follows:

```
Recursive function Head/tail Breaks:
    Rank the input data values from the biggest to the smallest;
    Compute the mean value of the data
    Break the data (around the mean) into the head and the tail;
    // the head for the data values greater than the mean
    // the tail for the data values less than the mean
    while (length(head)/length(data)<=40%):
        Head/tail Breaks(head);
End Function
```

Note that 40% is the threshold for the condition of whether to continue partitioning for the head. In other words, if the head percentage is greater than the set threshold, the function will stop. However, for many real-world data, this 40% threshold for every head can be relaxed to 40% on average for all the heads. This implies that, for some iterations, we can break up the 40% as long as, on average, the head percentage is equal to or less than 40%. The relaxed version of head/tail breaks is called head/tail breaks 2.0 (Jiang 2019b), while the above version is called head/tail breaks 1.0.

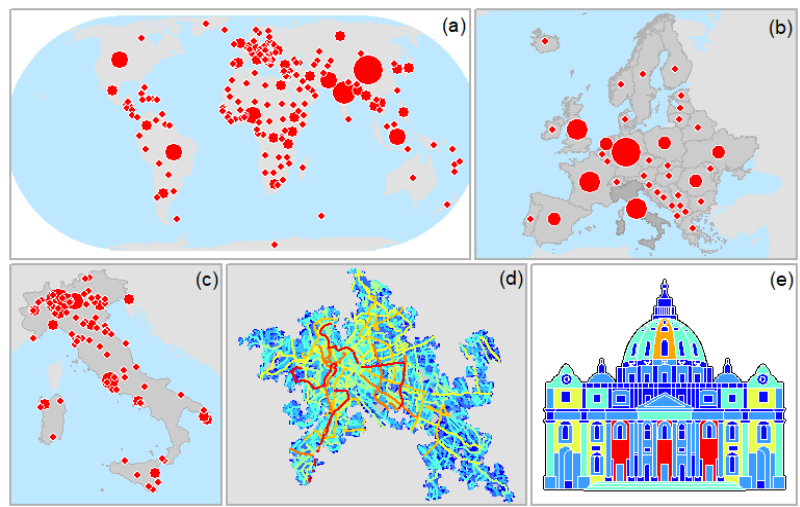

Figure 5: (Color online) The Earth's surface as a coherent whole, being a living structure
(Note: The living structure – scaling hierarchy of "far more smalls than larges" –
recurs at different levels of scale of the Earth's surface from (a) the global scale, (b) the European scale, (c) the country scale of Italy, (d) the city scale of Rome, and to (e) the building scale of St. Peter's Basilica. For panels (d) and (e), the blue indicates the least-connected or smallest, red indicates the most-connected or the largest, and other colors are between the smallest and largest.)

**4.2 The Earth's surface as a living structure from the globe to the building façade**
The Earth's surface is a living structure, seen from the global scale, down to the continent, to the country, to the city, and to the building façade, as shown in Figure 5. At every scale, there are "far more smalls than larges". For example, at both the global and continental scales, there are "far more small countries than large ones" in terms of their population (Figure 5a, 5b). At the country scale, there are "far more small cities than large ones" (Figure 5c). At the city scale, there are "far more less-connected streets (by cold colors) than well-connected ones (by warm colors)", with blues being the least-connected, reds being the most-connected, and other colors between the least- and most-connected. At the building scale, the façade of St. Peter's Basilica contains "far more small centers than large ones". It is important to note that, across the scales ranging from the globe to the city, there is no global symmetry, but there are full of local symmetries that make the Earth's surface beautiful or alive. There are "far more smalls than larges" globally, yet at each scale, things are "more or less similar". It is scaling law and Tobler's law that govern the Earth's surface as a living structure, being beautiful and alive.

Unlike many larger scales, building façades usually maintain their global symmetry, just as carpets must (Alexander 1993). However, for building plans, there is no need to retain the global symmetry, like the plan of the Alhambra (Alexander 2002–2005, Jiang 2015b), and the Eishin campus (Alexander et al. 2012, Guttmann et al. 2019). The Taj Mahal is undoubtedly a living structure, and it holds seven hierarchical levels of scale, some of which are illustrated in Figure 6. In this figure, we show different scales down to centimeters, but it can actually be shown down to millimeters, or the scale of an ornament. It should be noted that the Taj Mahal's facade is probably too symmetric or too restrict in terms of scaling ratio. The Taj Mahal is indeed living, but can become more living if some of the 15 properties – for example, alternating repetition – can be introduced, just as the Koch curve is indeed living, but it is less living than the coastline, as remarked in Section 2. Essentially, the Earth's surface in the wide range of $10^{-3}$ up to $10^7$ is a living structure. This is the general conclusion drawn from the case studies.

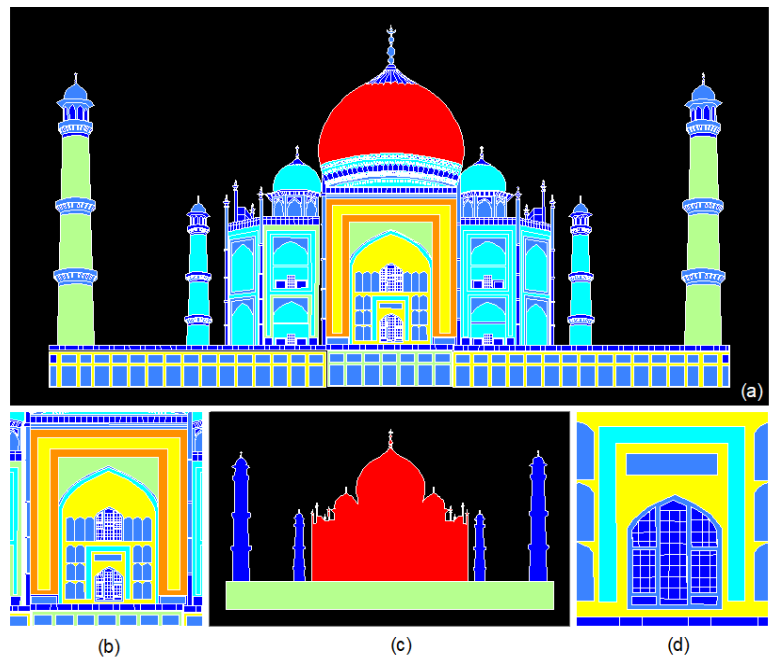

Figure 6: (Color online) The Taj Mahal as a living structure from the façade perspective
(Note: There are "far more small centers than larges" in the façade (a) with blue being the smallest, red being the largest, and other colors in between the smallest and largest. This is the essence of objective beauty or life. The scaling hierarchy of "far more smalls than larges" recurs six times; for example, (b), (c), and (d) each indicate an occurrence. The notion of "far more smalls than larges" is more powerful than the concept of self-similarity of fractal geometry (Mandelbrot 1983), because the former enables anyone – experts or laymen – to see fractals or living structures not only in nature, but also in what we build and make, such as buildings, art, and designs.)

Based on the Taj Mahal study, we can draw another (potentially controversial) conclusion that the Sydney Opera House is a less-living structure, for it clearly misses this kind of steep scaling hierarchy that we see in the Taj Mahal. Instead, the Sydney Opera House shows a very flat scaling hierarchy if any. The opera house may look a "good shape" on the surface, for it looks like a series of gleaming white sail-shaped shells. Seen from kilometers away, there are indeed "far more small shells than large ones", which occurs just once or at one scale rather than across scales, and monotonic repetition rather than "more or less similar" which recurs on each scale (Figure 7). The property of good shape does not necessarily imply any biological shapes, as seen on the surface (Thompson 1917), but something in far deep under the surface, the recurring scaling hierarchy of "far more smalls than larges". This is a misperception on good architecture by many of experts and the general public, just as the Golden Ratio is misunderstood as any shape with a ratio of ~1.618 (Salingaros 2018). To paraphrase Alexander, beauty is not skin-deep, but lies deep in the fine structure, or in the scaling hierarchy of "far more smalls than larges". Despite missing the scaling hierarchy, the opera house exhibits some of the 15 properties, so it cannot be said to be ugly when compared to many modernist buildings. For example, seen from kilometers away under the blue sky and white clouds, it looks really beautiful; Or seen from kilometers away under the background of the deadly boring tall buildings, it looks really living. Here we are talking about the property of not separateness. Therefore, living or less-living is relative rather than absolute, very much like the feeling of warmness.

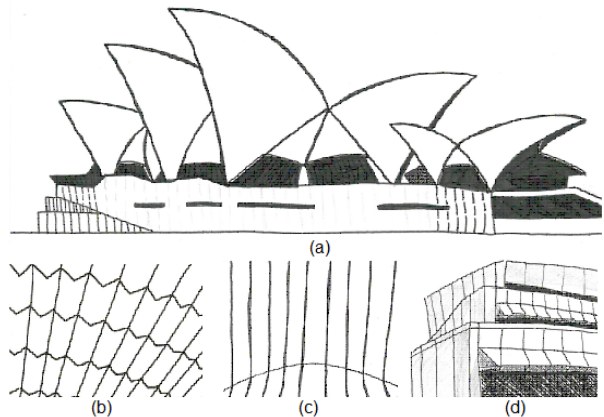

Figure 7: The Sydney Opera House as a less-living structure
(Note: The scaling hierarchy of "far more smalls than larges" occurs only once (a), and the opera house shows monotonic repetition at many levels of scale or on each scale (b), (c), (d). The drawing is by Celine Hedin.)

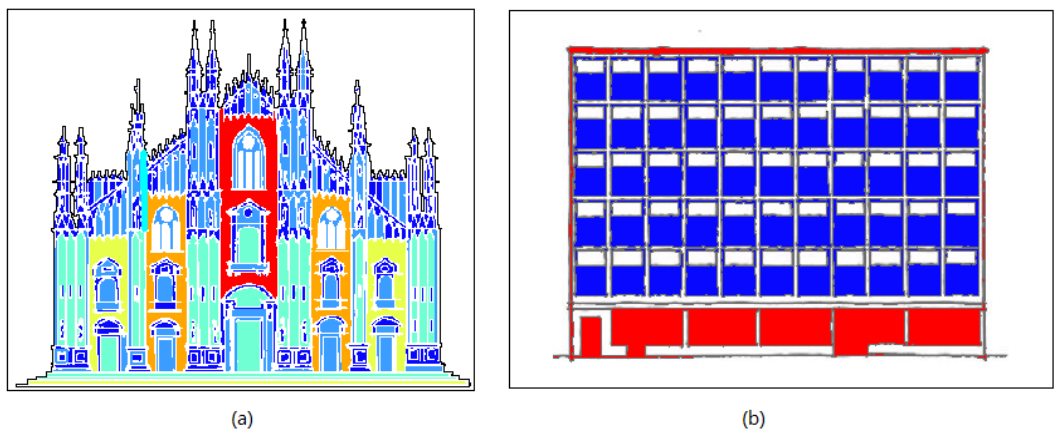

Figure 8: (Color online) Two building façades with different degrees of life or beauty
(Note: The façade on the left (a) has a steep scaling hierarchy with six levels of scale (the six colors), whereas the one on the right (b) has a very flat scaling hierarchy with only two levels of scale (the two colors). More importantly, all blue centers of the right façade are exactly the same without any variation. Beauty or life is determined (1) by the number of centers – the more centers, the more beautiful − and (2) by the scaling hierarchy – the more levels of scale, the more beautiful structurally.)
Source: This figure is created by the author based on the scanned black and white images from Kleineisel (1970).

**4.3 Comparison study on living structure**
This comparison study aims to prove that the modernist architecture and design usually fail to create living structure, whereas traditional buildings and designs are usually living structures. The study is constrained to building façades and all their identified centers, which are shown as individual polygons (Figure 8). The cathedral façade (Figure 8a) contains over 500 centers, whereas the modernist one (Figure 8b) contains only a bit over 50 centers. From these two numbers, we can judge that the left (Figure 8a) is more living than the right (Figure 8b). Secondly, the cathedral façade has six hierarchical levels indicated by the six colors (Figure 8a), whereas the modernist façade has only two levels represented by blue and red (Figure 8b). In addition, all of the blue pieces on the modernist building façade are exactly the same size, so look boring without any variation. There is little doubt that the left is more beautiful or living than the right.

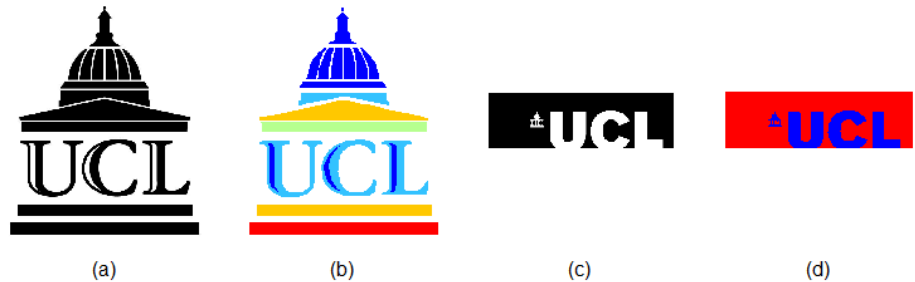

Figure 9: (Color online) Two UCL logos with different degrees of beauty or life
(Note: The old logo (the left, panel a and b) is more beautiful or more living than the new one (the right, panel c and d). This is because the old one has at least 19 centers with a scaling hierarchy of 5 by the five colors, whereas the new one has a maximum of six centers with a scaling hierarchy of only 2 by blue and red. Beauty or life is determined (1) by the number of centers – the more centers, the more beautiful − and (2) by the scaling hierarchy – the more levels of scale, the more beautiful structurally.)

The second comparison study applies to the two logos of University College London (UCL) (Figure 9). The new logo (http://tiny.cc/jpgxaz) was adopted in 2005, but it is far less-living than the old one (http://tiny.cc/81gxaz). Now let's examine the number of centers and the scaling hierarchy they hold. The old logo has at least 19 centers, which hold five hierarchical levels (Figure 9a and 9b), whereas the new logo has a maximum of six centers, which can be put at only two hierarchical levels: the five centers as the figure, and the one center as the ground (Figure 9c and 9d). In fact, two of the six centers are a bit too small to recognize when the logo is small enough. It is clear that the old logo is more living or more beautiful than the new one. Even by assessing the three letters U, C, L, we can conclude that the old one is more beautiful or more living than the new one, based on the fact that serif letters in general have more centers – thus more beautiful – than sans-serif ones. To this point, the more-living and less-living logos can really be said to a fact rather than an opinion.

## 4.4 Discussion on the case studies
The case studies demonstrate that living structure meets both the scaling law and Tobler's law, whereas dead structure violates these two laws. For example, the two bad designs – the modernist façade and logo – are considered to be dead structures, for they have only two hierarchical levels (indicated by red and blue). It should be noted that these two bad designs are, by far, not the worst. All those buildings labeled by such names as modernism, postmodernism, and deconstructionism belong to so-called dead structures or disorganized complexity (Jacobs 1961). In this connection. we found no better paper than the one by Salingaros (2014) that illustrates vividly complexity in architecture and design; essentially, living structure is organized complexity, while dead structure is disorganized complexity. Dead structure or disorganized complexity violates scaling law across scales or Tobler's law on each scale, and is not created by differentiation and adaptation through the 15 structural properties. Ironically, all these dead structures were mistakenly claimed to be inspired by new complexity science such as fractal geometry and chaos theory by Jencks (2002), which mistakenly confused fractals with fragmented and disorganized things. The world of architecture needs a new paradigm, built on the phenomenon of living structure (Grabow 1983). Although this paradigm was put forward a long time ago, mainstream architecture is still very much dominated by some wrong ways of thinking.

Structure with different degrees of scaling hierarchy can evoke different senses of feeling: *"To the extent they are alive, they let our inner forces loose, and set us free; but when they are dead they keep us locked in inner conflict"* (Alexander 1979). This feeling that sets us free or the correlation between living structure and human's selfness can be revealed by the mirror-of-the-self experiment (Alexander 2002–2005, Wu 2015, Rofé 2016). Two things or their pictures are put side by side, and the human subject is asked to pick the one that best reflects oneself. Most of time and for most people, the one with a higher degree of wholeness was picked with a high degree of agreement. However, unlike many human experiments in psychology for seeking an inter-subjective agreement, the mirror-of-the-self experiment seeks the true feeling of wholeness. This kind of experiment is able to capture the genuine or objective feeling of wholeness inside the human beings, if questions about inner nature of wholeness are asked carefully and in the right way. In other words, a human being is a reliable measuring instrument for wholeness, leading to reliable and shared results. However, the question is, why can living structure trigger the feeling of beauty or life in human minds?

Despite the high correlation between living structure and the selfness, it is impossible – under the current mode of Cartesian thought – to set up the causation between the living structure and our inner world; that is, how the living physical world is reflected as our feeling of beauty and life. We know from statistics that correlation does not mean causation, and it has to be established scientifically. To address the question of why the feeling of life and beauty can be triggered by living structure, Alexander put forward an I-hypothesis or a new cosmology (Alexander 2002–2005, Book 4, p. 23), *"one in which the idea of great art is possible – even necessary – as something which connects us to the universe, something which can provide a proper underpinning for the art of building"*. The next section will concentrate on the I-hypothesis and new cosmology.

**5. The inner meanings of living structure through the I-hypothesis**
The phenomenon of living structure implies that the real world is a living world, to which we human beings belong. This section further discusses the I-hypothesis, the new cosmology, and how to make better sense of living structure in our inner world.

**5.1 The I-hypothesis and how it comes**
The I-hypothesis states that there is, physically in the universe, and underlying all space and matter – at every point of space and matter – a single underlying substance that shall be simply called "I". It can be called or expressed in a variety of ways: the concept of "I", the universal "I", the luminous ground, the blazing one, the I-substance, the "I", the Self, the heaven, the spirit, the soul, the domain of "I", the ground of "I", the eternal "I", the plenum of "I", something luminous, the underlying "I", the concept of beings, and even God. In this paper, I do not refer to it as God but instead as the "I", hence the I-hypothesis. The I-hypothesis goes beyond the mechanistic world picture, under which the human inner world is separated or disconnected from the physical world. The I-hypothesis implies an organic world view, under which the human inner world is tightly united with the physical world through the "I". This is difficult for our minds to understand, because we have become used to the mechanistic world picture. The mechanistic world view is superficial, and could be simply wrong (Whitehead 1938). We have to get out of the mode of thought to which we have become accustomed, just as Alexander did, in order to make better sense of the inner meaning of living structure. This philosophical and visionary part of living structure is metaphysical, and remains highly mysterious, and probably can never be verified.

The following quote shows how Alexander (2002–2005, Volume 4, p. 136) himself struggled with the I-hypothesis:

*"When thinking as a scientist, it must of course be this question of truth which occupies one's mind. It is for this reason that I have kept records, and written down my observations, for the last thirty years, as carefully as possible. As a result of my observations, and as a result of my experiences in the field − as an architect building buildings, as a craftsman making things, as a planner laying out buildings and precincts and seeing them come to life − I have gradually become convinced that this theory* [the I-

hypothesis, note by this paper's author], *or at least something very much like it, is indeed likely to be true. In short, as a scientist, I have gradually come to the belief that the I must be real. And as an architect, I have also become convinced that the I is certainly real in buildings, and must necessarily play a fundamental role in architecture."*

This I-hypothesis or a version of it first came to Alexander's mind while he was studying Turkish carpets. As he wrote at the very beginning of the book:

*"A carpet is a picture of God. That is the essential fact, fundamental to the people who produced the carpets, and fundamental to any proper understanding of these carpets.*

*This does not mean, in Anglo-Western terms, that a carpet is a picture of a man with a long white beard. God, the all seeing, everlasting stuff, is the target of Sufism - as it is of all the mystical religions. In modern language we might also call it ultimate oneness of everything. The Sufis, who wove most of these carpets, tried to reach union with God. And, in doing it, in contemplating this God, the carpet actually tries, itself, to be a picture of the all seeing everlasting stuff. We may also call it the infinite domain or pearl-stuff."* (Alexander 1993)

**5.2 The new organic cosmology**
From the I-hypothesis, Alexander conceived and conceptualized a new world view or new cosmology, in which the outer physical world and the inner emotional world are united as one, in one coherent world picture. Our individual inner selves and any living structures in space and matter are imbued with the "I". The density of the "I" is not uniformly penetrated in space and matter, and it depends on the degree of wholeness or living structure. In other words, the more living a center, the greater the possibility that the center reveals the "I". It is essentially a non-material view of space and matter, and is therefore hard for our mechanistic mindsets to accept. If it were accepted, it would make good sense in terms of explaining why the feeling of beauty or life – or our inner world – can be triggered by the living structure of the physical outer world.

The new cosmology put forward by Alexander has its philosophical and religious roots. For example, Whitehead (1920) believed that we cannot have a proper grasp of the universe and our place in it until two worlds – the physical world and our experienced inner world, which are called "bifurcation of nature" – can be united in a single picture. This is exactly what Alexander did to get these two worlds united through the hypothesized "I", and based on the powerful notion of living structure. Alexander felt strongly that to build great buildings, and make great arts, one must keep the creation or design as a worship to the "I" or God. According to Alexander, the weavers wanted to be united with God through the carpet as a living center. *Places for the Soul* is a 30-minute film on the work of Christopher Alexander by an independent filmmaker, which highlighted Alexander's approach to building. In it, Alexander remarked that his approach to architecture was "*to make God appear in the middle of field*". In the first paragraph of the book on carpets (Alexander 1993), as mentioned above, he began with the sentence "*A carpet is a picture of God*", while in his essay, *The long path that leads from the making of our world to God* (Alexander 2007a), he made an explicit link between the built environment and God. All these point to the fact that his thoughts are pretty religious. If we thought anything we make or create is to sense and to see the "I", would our daily life not become more meaningful?

It should be noted that the new organic world picture is not about abandoning the current mechanistic world view, but about the hypothesized "I" to make the current world picture complete and integrated. Human beings are not separated from the physical world, as is currently conceived, but united with the outer world through the ubiquitous hypothesized "I". This new world makes better sense as to why our consciousness comes from the human brain, which is the most living center in the body, and enables us to sense or see the ground "I".

**5.3 Making sense of living structure in our inner world**
Let's see how the I-hypothesis makes better sense of living structure in our inner worlds than the purely psychological view. According to Alexander (2002−2005), space is neither lifeless nor neutral, but a living structure capable of being more living or less living. This living space view includes everything in the physical world, as large as the universe ($10^{27}$), as small as the Planck size ($10^{-35}$), and sizes in between the largest and the smallest. This living space view is very much like extending the picture of Figure 5 at its two ends: beyond the Earth toward the entire universe, and further down from the building façade toward the smallest. In this living space, according to the I-hypothesis, the "I" exists everywhere. Therefore, any human being is part of this living space, and he – the human body – is also fused or tunneled with the "I". Like any space, the human body is not uniform in terms of living centers; some parts are more living and some parts are less so. For example, the human brain is the most living center in the human body. This explains why we have consciousness or why we have the feeling of beauty or life, and why great architecture and works of art are religious, and why ancient carpets are religious. This is because the human brain is the most living center, through which one can sense or see the "I". In other words, the living center acts like a window, through which one sees the "I": the more living a center is, the higher the possibility one sees the "I" through the center. Whether the I-hypothesis is true may never be verified. However, this physical or metaphysical view of the universe seems better than the psychological explanation about beauty and life.

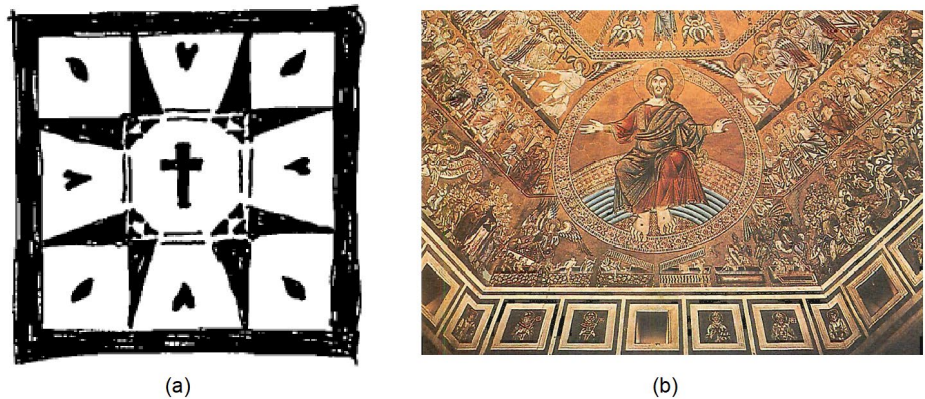

Figure 10: (Color online) Living structure enabling us to see the luminous ground
(Note: (a) A small humble panel made of marble strips inlaid in the west wall, and (b) gold and mosaic ceiling, of the Baptistery, Florence, 11th century (Alexander 2002–2005).)

A living structure – or a living façade in particular – is full of hundreds or thousands of living centers, which Alexander (1993) called "beings". A being is the most beautiful or most living center in a living structure. It is autonomous, capable of uplifting the human spirit. For example, the front gate of the Taj Mahal façade is such a being or being-like center (Figure 6b). One could also say that the upper dome is being-like because it contains many elaborate and intricate sub-structures. As Alexander (2002–2005) gazed at the humble yet beautiful tile (Figure 10a), he felt clearly as though he was looking through to heaven. This expression of his own feeling may seem far-fetched or romantic under the current mechanistic mode of thought, but it should be understood as something literal under the new world view, a non-material view of space and matter. To mimic how Alexander expressed his own feeling, we could say that by gazing at any picture of Figures 5, 6 and 10, our spirit is uplifted. Through contemplation, we develop a deep sense of feeling that we are part of Earth, part of the building, or part of the city, and eventually part of the universe. In the same fashion, any picture of Figure 6 enables us to see the "I" or the luminous ground, or we can become united with the luminous ground and our spirit is uplifted. In a secular tone or in the current mechanistic mode of thought, we feel at ease emotionally, as our personal feeling, or our true feeling.

**6. Conclusion**
This paper is intended to defend living structure as a physical phenomenon and mathematical concept for people to understand objective or structural nature of beauty, or to setup a dialogue with those who are skeptical about Alexander's profound design thoughts. Living structure is a mathematical structure of physical space, which is able to reflect in our minds psychologically: the more living the structure is, the more beautiful one feels. By drawing evidence from Alexander's work and through our own case studies, this paper has shown that beauty is essentially objective or structural. In other words, beauty exists in the deep structure of details, or in the scaling hierarchy of "far more smalls than larges". Beauty and ugliness can be clearly defined by scaling law; that is, a structure with a flat scaling hierarchy – with maximum two levels of scale only – is objectively considered to be ugly, whereas a structure with a steep scaling hierarchy – with at least three levels of scale – is objectively considered to be beautiful.

Armed with the kind of simple analysis on scaling hierarchy, people can understand why beautiful buildings are beautiful, and why ugly buildings are ugly. Simply put, a building is beautiful because of its steep scaling hierarchy, or ugly because of its flat scaling hierarchy. By claiming objective or structural beauty, our intention is not to deny idiosyncratic aspects of beauty, which account for only a small proportion of our feeling. This dominance of the objective over the subjective can be compared to any statistical regularity with a majority of agreement, such as an r square value of 0.75 instead of 1.0. In addition to the scaling hierarchy or scaling law, Tobler's law plays an important role in the objective or structural beauty as well. As one of the two laws of living structure, Tobler's law – or the notion of "more or less similar" – recurs on each level of scale. The true meaning of "more or less similar" is neither "completely same" nor "completely unique", but something between the same and the unique. These two complementary laws work together, governing living structures, with the scaling law being primary, and Tobler's law being secondary.

The I-hypothesis is a powerful concept that makes better sense of the inner meaning of living structure than purely psychological or cognitive explanations. The third view of space provides a fresh look at our surroundings, whereby everything is a living structure, and should become more living or more beautiful through our daily makings. The new cosmology solves the problem of the bifurcation of nature (Whitehead 1920), for we human beings are not separate from, but are part of the universe, thus making our daily lives more meaningful. Human beings can be uplifted by good space, reflected by good architecture, and eventually united and re-united to the hypothesized "I". To end this paper, we would like to claim that living structure may actually be the *"bead game conjecture"* (Alexander 1968, cited from Gabriel 1998, and Grabow 1983), a mechanism that unites all structures or forms in mathematics, science, art, philosophy, and religion. This claim requires further research on living structure from these multiple disciplines or contexts, and implies that living structure is not a dogma, and can instead be further discussed, argued and even challenged.

**Acknowledgement**
XXXXXXXXXX